Uncorrelated Volatile Behavior During the 2011 Apparition of Comet C/2009 P1 Garradd

Lori M. Feaga[1], Michael F. A'Hearn[1], Tony L. Farnham[1], Dennis Bodewits[1], Jessica M. Sunshine[1], Alan M. Gersch[1], Silvia Protopapa[1], Bin Yang[2], Michal Drahus[3,4], David G. Schleicher[5]

[1]Department of Astronomy, University of Maryland, College Park, MD 20742, USA
Corresponding author email: feaga@astro.umd.edu
[2]Institute for Astronomy, University of Hawaii, Honolulu, HI 96822, USA
[3,4]Division of Physics, Mathematics & Astronomy, California Institute of Technology, Pasadena, CA 91125, USA; National Radio Astronomy Observatory, Charlottesville, VA 22903, USA
[5]Lowell Observatory, Flagstaff, AZ 86001, USA

Short Title:  Uncorrelated Volatile Behavior of Comet C/2009 P1 Garradd

**ABSTRACT:**

The High Resolution Instrument Infrared Spectrometer (HRI-IR) onboard the Deep Impact Flyby spacecraft detected $H_2O$, $CO_2$, and CO in the coma of the dynamically young Oort cloud comet C/2009 P1 (Garradd) post-perihelion at a heliocentric distance of 2 AU. Production rates were derived for the parent volatiles, $Q_{H2O}$ = 4.6 ± 0.8 × $10^{28}$, $Q_{CO2}$ = 3.9 ± 0.7 × $10^{27}$, and $Q_{CO}$ = 2.9 ± 0.8 × $10^{28}$ molecules $s^{-1}$, and are consistent with the trends seen by other observers and within the error bars of measurements acquired during a similar time period. When compiled with other observations of Garradd's dominant volatiles, unexpected behavior was seen in the release of CO. Garradd's $H_2O$ outgassing, increasing and peaking pre-perihelion and then steadily decreasing, is more typical than that of CO, which monotonically increased throughout the entire apparition. Due to the temporal asymmetry in volatile release, Garradd exhibited the highest CO to $H_2O$ abundance ratio ever observed for any comet inside the water snow line at ~60% during the HRI-IR observations. Also, the HRI-IR made the only direct measurement of $CO_2$, giving a typical cometary abundance ratio of $CO_2$ to $H_2O$ of 8% but, with only one measurement, no sense of how it varied with orbital position.

**Subject Keywords:** comets: general, comets: individual (C/2009 P1 Garradd), infrared: planetary systems, techniques: spectroscopic

1. **INTRODUCTION**

Since comets are primarily composed of the primitive ices and grains left over from the formation of the Solar System 4.5 Gyr ago, a study and comparison of their volatile composition helps to characterize the primordial material, conditions, and degree of mixing within the solar nebula. When a comet approaches the Sun, its surface temperature increases and the solar heat wave penetrates to subsurface reservoirs where preserved volatile materials sublime and form jets and a coma. Therefore, temporal studies of comets throughout their orbits, rather than single snapshots, have the potential to give more detailed information on the heterogeneity of the volatiles stored in the nuclei, abundance and depletion of primitive volatiles, and possible insight into the evolutionary processes that have occurred on the comet.

To date, great diversity in composition and activity has been seen among comets and even within groups of comets (i.e. Jupiter family and long-period comets), which suggests that there was significant mixing and overlap in the cometary formation regions in the solar nebula (A'Hearn et al. 1995, Biver et al. 2002, Mumma et al. 2003, Bockelée-Morvan et al. 2004, A'Hearn et al. 2005, A'Hearn et al. 2011, Mumma & Charnley 2011, Ootsubo et al. 2012, A'Hearn et al. 2012) and that the release of volatiles is complex and not well understood. Recent dynamical models can explain such a high degree of mixing during formation of the current cometary reservoirs (Gomes, et al 2005, Tsiganis et al. 2005, Dodson-Robinson et al. 2009, Levison et al. 2011, Brasser & Morbidelli 2013). For several of these comets, only a snapshot in time is available for studying the volatile release, which (absent a long-term trend) may not adequately represent the bulk comet composition. To improve understanding of how comets work as well as initial conditions and mechanisms at work during their formation, the sample size of comets for which the temporal behavior of parent volatiles has been studied must be increased.

The dominant parent volatile species in comets are, in order of increasing volatility, $H_2O$, $CO_2$ and CO. All three of these volatiles fluoresce in the infrared due to ro-vibrational transitions that can be observed (Yamamoto 1982, Crovisier & Encrenaz 1983, Weaver & Mumma 1984). $H_2O$ can be observed from ground-based telescopes, as can CO when there is a substantial geocentric Doppler shift or at mm wavelengths. Both species can also be observed with space-based observatories. $CO_2$ cannot be observed from platforms within Earth's atmosphere because the absorption bands of telluric $CO_2$ are opaque even at SOFIA's altitudes. This reduces the number of available measurements of the $CO_2$ abundance in comets. Moreover, even fewer comets have measurements of both $CO_2$ and CO (Bockelée-Morvan et al. 2004, Mumma & Charnley 2011, Ootsubo et al. 2012, A'Hearn et al. 2012) making it more difficult to parameterize the volatile distribution. For comets studied inside of 2.5 AU, abundances of CO and $CO_2$ relative to $H_2O$ vary between 0-35% (Ootsubo et al. 2012, A'Hearn et al. 2012).

In the late 1990's, the extremely bright and dynamically young Oort cloud comet C/1995 O1 (Hale–Bopp) gave cometary scientists a rare opportunity to monitor the outgassing behavior of several volatile species, both pre- and post-perihelion, over a large range of heliocentric distances and across the spectrum (Biver et al. 1997, Feldman 1997, Mumma et al. 2003). In general, the extensive temporal coverage showed that the production rates of all volatile species increased with decreasing heliocentric distance, peaked near perihelion, and decreased again post-perihelion with increasing heliocentric distance. In 2009, another exceptional Oort cloud comet, C/2009 P1 (Garradd), was discovered (McNaught and Garradd 2009) which again afforded the cometary community the ability to quantify another young comet's volatile release with time. Early on, Garradd was determined to be enriched in carbon monoxide (Paganini et al. 2012) and multiple sets of observations spanning 1.5 years around Garradd's 2011 Dec 23 perihelion passage ($R_h$ = 1.55 AU) demonstrated that the volatile behavior at Garradd was very different from that of Hale-Bopp. Garradd's water production rates peaked well before perihelion (Combi et al. 2013, Bodewits et al. 2013, in preparation, Farnham et al. 2013, in preparation). In contrast, as presented below, the carbon monoxide production continued to increase through perihelion and beyond.

After successful flybys of comets 9P/Tempel 1 and 103P/Hartley 2 (A'Hearn et al. 2005, A'Hearn et al. 2011), the Deep Impact Flyby (DIF) spacecraft and its fully operational instrument suite were used to observe comet Garradd post-perihelion from a distance of 1.4 AU. The High Resolution Instrument Infrared Spectrometer (HRI-IR; Hampton et al. 2005) was the only operational and available instrument with the ability to directly and simultaneously detect $H_2O$, $CO_2$ and CO in Garradd, augmenting the ground-based measurements of $H_2O$ and CO, and reducing systematic errors generated from inferring production rates from daughter or granddaughter species (e.g., OH, H for $H_2O$ and O for CO and $CO_2$). The DIF platform was also able to provide continuous temporal coverage of a dedicated target without the day/night scheduling issues of many other telescopic assets, thus further reducing systematic errors in the derivation of absolute abundance ratios resulting from piecing together data across hours or nights of observing or from different telescopes. The DIF infrared observations presented in this paper provide the only direct measurement of $CO_2$ for Garradd in addition to the latest post-perihelion CO measurement to date and contribution of a direct measurement to the wealth of water production rates (derived from both direct and indirect measurements of $H_2O$) for the comet (Combi et al. 2013, DiSanti et al. 2013, Farnham et al. 2013, in preparation, Bodewits et al. 2013, in preparation, Bodewits et al. 2012, Feldman et al. 2012, Paganini et al. 2012, Villanueva et al. 2012b, Bockelée-Morvan et al. 2012). DIF visible camera observations will be discussed by Farnham et al. (2013, in preparation).

## 2. OBSERVATIONS

The DIF spacecraft is equipped with medium- and high-resolution CCD imagers (MRI and HRI) and an infrared spectrometer (HRI-IR), with wavelength coverage from 1.05 to 4.83 μm (Hampton et al. 2005). Ro-vibrational bands of $H_2O$ ($v_3$ centered near 2.67 μm), $CO_2$ ($v_3$ centered near 4.26 μm), and CO (v(1,0) centered near 4.67 μm) lie within this spectral range. The minimum resolving power of the double prism spectrometer ($\lambda/\delta\lambda$) is ~200 at ~2.6 μm, with higher resolving power (>400) achieved at the ends of the spectrum. In its default operational mode, HRI-IR physical pixels are binned and have an instantaneous field of view (IFOV) of 10 μrad, with 256 pixels spatially across the frame and 512 spectral channels. Because warm targets, specifically the nucleus of comet 9P/Tempel 1 (the target of the DI primary mission), produce high thermal signals at longer wavelengths, a spectral attenuator (anti-saturation filter) covers the middle third of the slit, suppressing the strong thermal contribution above 4 μm (Klaasen et al. 2008, Klaasen et al. 2013) for targets at small heliocentric distances. For observations of unresolved cometary targets and coma studies, the slit is positioned such that the signal falls on the unfiltered regions of the detector.

Post-perihelion MRI visible observations of Garradd spanned 20 February 2012 through 9 April 2012 when Garradd traveled from 1.7 to 2.1 AU and varied between 9[th] and 10[th] total visual magnitude as seen from the Earth and a little fainter as seen from the DIF. From these data, dust and gas production rates (OH, CN) were estimated, a lightcurve was extracted, and a rotation period of 10.4 hr was determined for the comet (Bodewits et al. 2012, Farnham et al. 2013, in preparation). While the DIF visible camera monitored Garradd for 1.5 months, the observing window for the HRI-IR spectrometer was much more limited due to solar elongation constraints.

HRI-IR scans across the unresolved nucleus and coma of Garradd, at ~35° solar phase angle, were collected every fifteen minutes for sixteen hours on each of two days, 26 Mar 2012 and 2 Apr 2012. The data will be archived in the Planetary Data System (http://pds.jpl.nasa.gov). The observing parameters can be found in Tables 1 and 2. Each scan included 17 frames collected in full-slit binned mode (BINFF, Hampton et al. 2005) with the spectrometer slit aligned with the Sun-anti-Sun line. The individual 12-second frames of a given scan were acquired in succession while rotating the spacecraft perpendicular to the slit (Figure 1A). Typically, HRI-IR scan rates are set to move 1 slit width (10 μrad) per HRI-IR frame. However, in order to account for estimated spacecraft pointing uncertainties for Garradd, the scan rate was increased to integrate over two slit widths (IFOV 20 μrad) in each exposed frame. The resulting spectra (e.g., Figure 1B) are thus averaged over a solid angle of 10 μrad along the slit by 20 μrad in the scan direction (2" × 4" or 2,105 km × 4,210 km projected at the comet). Each 17-frame scan was converted into a two dimensional spatial image along the scan and slit directions to facilitate examination of compositional asymmetries as described below.

Based on the improved approach for the Hartley 2 flyby (Klaasen et al. 2013), the HRI-IR data were calibrated using the following technique. A pixel-dependent 4[th] order polynomial was used to linearize the response of each pixel in a frame, followed by subtraction of a dark frame obtained from averaging the final two frames of the 17-frame scan. A flat field correction and an absolute flux calibration were then applied to each frame resulting in an average surface brightness in each pixel of each frame with units of radiance (W m$^{-2}$ sr$^{-1}$ μm$^{-1}$). A full description of the improved HRI-IR calibration pipeline developed for the Hartley 2 encounter can be found in Klaasen et al. (2013). The applicable calibration files, a description of the pipeline processing, and limitations in the calibration, were archived in the Planetary Data System (http://pds.jpl.nasa.gov, McLaughlin et al. 2013).

At Garradd's distance from the Sun (2 AU) and distance from the DIF spacecraft (1.4 AU), the cometary continuum signal was not detectable in individual HRI-IR frames. Therefore, the location of Garradd's nucleus in the HRI-IR scans (along the slit and across the scan) was determined by locating Garradd in MRI visible context images (10 μrad IFOV) acquired immediately after each HRI-IR scan and correcting for the known boresight offset between the instruments and spacecraft motion during the scan. Although the location of the comet in the MRI image was measured to the sub-pixel level, HRI-IR data were only aligned at the pixel level. After spatially co-aligning the images, a single data set (time-averaged over both days) was generated to maximize the signal to noise. Once stacked, we calculated the mean value, excluding data points lying more than 2.5 σ from the median value when calculating the mean (resistant mean), for each pixel in the 128 scans. This resulted in a single spatial-spectral image of Garradd from 1.05 to 4.83 μm.

In order to remove any remaining background sky signal, a sky spectrum, extracted as a resistant mean of 35 pixels from the opposite side of the slit from Garradd's position (Figure 1B), was subtracted. An example of the resulting spatial-spectral image, used in subsequent analyses, is shown in Figure 1B. Non-negligible sources of error in the brightness image stem from random noise in the low signal data (ranging from 5-30% of the emission band intensities), a systematic uncertainty in the absolute flux calibration (at the 10% level; Klaasen et al. 2013), and systematic uncertainty from averaging the data set over a full rotational period (<10%).

## 3. ANALYSIS

### 3.1 Column Density

Details of the analysis, from extracting a spectrum to deriving production rates, are described below, using the Garradd spectrum with the highest SNR as a representative example. The highest SNR spectrum was extracted for a 5-pixel (100 μrad in the scan direction) by 9-pixel (90 μrad in the slit direction) rectangular aperture that included and was centered on the unresolved nucleus (Figure 2). This aperture projects to a 21,050 km by 18,945 km area at Garradd and will hereafter be referred to as the large aperture. This large aperture is nearly square and incorporates the entire detectable coma signal without integrating excess noise from the background.

### 3.1.1 Molecular Bands

Visual inspection of the extracted spectrum (Figure 2) shows obvious emission bands, but very little underlying continuum from reflected solar light and thermal emission by the nucleus or coma particles. The $\nu_3$ (1,0) bands of $H_2O$ and $CO_2$ are very prominent, similar to those seen in DIF spectra of Tempel 1 and Hartley 2 (Feaga et al. 2007a; A'Hearn et al. 2011). Detection of the CO $\nu(1,0)$ band is also evident at the long-wavelength end of the spectrum. In the post-impact ejecta cloud of Tempel 1, the CO signal was superposed on a very strong thermal continuum and thus difficult to analyze (Feaga et. al. 2007b), but the CO signal in Garradd is more easily interpreted because the thermal continuum was absent. The SNR for these parent molecular bands is ~10 for $H_2O$, ~8 for $CO_2$, and ~5 for CO in the representative spectrum. In addition to these parent molecules, an unidentified band at 2.8 µm, longward of the water, was detected and will be discussed later in the paper. Average band surface brightnesses were measured for $H_2O$, $CO_2$, and CO in the extracted spectra between 2.56-2.75, 4.18-4.33, and 4.58-4.75 µm respectively (Table 3). The wavelength ranges over which the bands were integrated were chosen to match both the features in the data as well as modeled band widths and are denoted with vertical dotted lines in Figure 2.

### 3.1.2 Comparison to 103P/Hartley 2

In Figure 3, the large aperture spectrum of Garradd is compared with a similarly constructed spectrum of 103P/Hartley 2, a comet for which the CO production rate was very low (0.15%–0.45%, Weaver et al. 2011). For the representative Hartley 2 spectrum, the DIF spacecraft was much closer to the comet and a slight, but non-negligible, thermal continuum was present. A fit to the thermal continuum was made and then subtracted before the residual spectrum was quantitatively analyzed. Unlike Garradd, CO was not detected in the HRI-IR Hartley 2 data.

### 3.1.3 Column Density Derivation

Column densities for the parent volatiles were derived from integrated band fluxes as follows. The average flux in a band was calculated by multiplying the average surface brightness of the band by the solid angle of a pixel. Assuming a spherically symmetric coma, fluorescent pumping, and a solar fluorescent g-factor that incorporates the optical depth of the band (Table 3), the following equation defines the average column density of a given species:

$$\mathcal{N}_{species} = 4\pi \, (d^2/s^2) \, F_{band} \, (\lambda_{band}/hc) \, (R_h^2/g_{band}) \qquad \text{Equation 1}$$

where $\mathcal{N}_{species}$ is the column density of gas measured [$cm^{-2}$], ($s^2/d^2$) is equal to the solid angle of an HRI-IR pixel, d is the spacecraft to comet distance [cm], $s^2$ is the area of a pixel at Garradd [$cm^2$], $F_{band}$ is the average flux from the gas emission band [erg $cm^{-2}$ $s^{-1}$], $\lambda_{band}$ is the central wavelength of the emission band [cm], h and c are Planck's constant and the speed of light [cgs units], $R_h$ is the heliocentric distance of the comet in AU, and $g_{band}$ is the effective fluorescent g-factor of the band at $R_h$ = 1

AU. For the spectrum in Figure 2, the column densities measured for $H_2O$, $CO_2$, and CO are $4.1 \pm 0.7 \times 10^{14}$, $3.5 \pm 0.7 \times 10^{13}$, and $2.6 \pm 0.7 \times 10^{14}$ cm$^{-2}$, respectively. A discussion of optical depth effects and the effective g-factors is presented in Section 3.2.2.

## 3.2 Varying Aperture Sizes and Opacity

Time-averaged spectra of Garradd were extracted for various aperture sizes, compared, and examined to determine the degree of opacity.

### 3.2.1 Apertures and Annuli
Spectra were extracted from various smaller rectangular apertures and rectangular annuli (region bounded by two concentric rectangles). They were then compared with the spectrum already presented in Figure 2. With fewer pixels to derive an average surface brightness, the smaller apertures are in general noisier than the spectrum presented in Figure 2. The smallest rectangular aperture extracted, hereafter referred to as the small aperture, was for the single central pixel that includes the unresolved nucleus of Garradd and subtends an area of 4,210 km (scan direction) by 2,105 km (slit direction) on the sky (Figure 4a). The parent molecular bands exhibit a SNR of ~5 for $H_2O$, ~4 for $CO_2$, and ~3 for CO in this spectrum. An intermediate sized aperture was also studied. The spectrum was extracted for a 3-pixel (scan direction) by 5-pixel (slit direction) rectangular aperture (12,630 km × 10,525 km), again centered on the nucleus (Figure 4b) with SNR closer to that of the large aperture spectrum from Figure 2. Qualitatively, emission bands from $H_2O$, $CO_2$, and CO are present in the spectra from all three apertures and are dominated by the water emission. Taking into account the noise, the relative intensities of the branches and overall shape of the $H_2O$ band are similar in all three spectra (Figures 2 and 4). $CO_2$ does not exhibit much relative change while CO appears to change from a double peak feature in the small aperture to a single peak in the large aperture. Spectra were also studied for rectangular annuli (Figure 5) by extracting the measured flux from successive rectangular frames corresponding to the regions between the boundaries of each of the rectangular apertures described above (14 pixels with an effective radius of 4,100 km and 30 pixels with an effective radius of 8,900 km). Column densities calculated from these extracted spectra are listed in Table 3.

### 3.2.2 Optical Depth
When optical depth is an issue, the relationship between band flux and column density in Equation 1 is no longer linear, and the derived column densities assuming optically thin conditions would result in lower limits. Complex radiative transfer equations need to be solved to most accurately account for the opacity of the coma. For the HRI-IR observations, the aperture-averaged spectra were modeled using the spherical adaption of the Coupled Escape Probability radiative transfer technique of Gersch & A'Hearn (2013, submitted). Production rates were initially calculated using optically thin g-factors for the large aperture ($g_{H2O}$ = 3.1 × 10$^{-4}$ s$^{-1}$ (Crovisier 1984a,b); $g_{CO2}$ = 2.8 × $10^{-3}$ $s^{-1}$ (Crovisier & Encrenaz 1983); $g_{CO}$ = 2.4 × $10^{-4}$ $s^{-1}$ (Chin & Weaver 1984)), where opacity was assumed to have little or no effect. These, together with a gas expansion velocity, a gas temperature, an $H_2O$ ortho-para ratio, and the size of the nucleus were supplied as inputs to the model. A 3 km radius was assumed for Garradd. The ortho-para ratio was set to 3:1, consistent with values found in some comets (Bockelée-Morvan et al. 2004, Mumma & Charnley 2011). For simplicity, we assumed a constant gas velocity (5 × $10^4$ cm $s^{-1}$) and constant temperature in the model throughout a symmetric coma. After testing various temperatures, the best-fit model used an input temperature of 40 K, consistent with rotational temperatures measured for several species in Garradd's coma by Villanueva et al. (2012b). The output of the model gave band shape, total band flux and effective g-factors to compensate for the opacity among other details. After an iterative process of matching these modeled parameters to the integrated band flux in the data, a revised production rate was established using these effective g-factors (Table 3). The model is overplotted on the small and large aperture data in Figure 6. For the HRI-IR Garradd data, the model predicted that the coma transitioned from optically thick to thin, i.e., had an effective g-factor ≥ 90% of the optically thin value, beyond 10,000 km from the nucleus for all species. It was also determined from the model that using optically thin g-factors to calculate column densities beyond 10,000 km from the nucleus would underestimate the actual values by less than 10% (Figure 7).

Figure 8 demonstrates the effects of optical depth on the measured surface brightnesses for Garradd, and thus the inaccuracy of the derived column densities if assuming optically thin conditions. If Garradd's coma were optically thin and followed a 1/r profile with constant gas outflow velocities, the average surface brightness in each annulus multiplied by the effective radius of the annulus would not change with distance from the nucleus. Consistent with the model predictions, the integrated surface brightness in the central pixel is lower than that in the intermediate and large rectangular annuli, meaning that optical depth effects are most significant for the innermost coma. However, all three distances are affected by differing degrees of optical depth for each molecule. Therefore, the modeled optically thick g-factors (Table 3) are an improvement over no opacity correction and are used in the column density calculations. The large annulus shows a decrease in the product of average surface brightness and distance from the nucleus for $H_2O$ and $CO_2$ rather than an increase (if optically thick) or approaching a straight line (if optically thin), suggesting that a constant velocity should not be assumed for Garradd within 10,000 km of the nucleus as has been done. Combi (2002) presented dynamical models of variable gas velocities in the coma due to collisional processes. At 2 AU, Combi's model for Hale-Bopp predicts that the gas velocity from ~1,000 km to ~10,000 km will increase by 15-20%. The HRI-IR analysis confirms that a velocity increase of this magnitude would be consistent with the data.

In an aperture of comparable size with the intermediate and large HRI-IR apertures, the MRI continuum and narrowband filter data show distinct arcs in brightness with only slight minimum to maximum variation throughout a rotation period (1% of the brightness in the continuum, 4% in CN and 6% in OH; Bodewits et al. 2012, Farnham et al. 2013, in prep).  Even though there are no simultaneous MRI OH measurements with the HRI-IR scans, the amplitude of variation for the contemporaneous CN is the same fraction during the HRI-IR scan observations as it was a month earlier when OH data were acquired, so it is assumed that the percent variability of OH does not change either.  Therefore, we infer that the deviation in $H_2O$ and $CO_2$ surface brightness between the intermediate and large annuli (~20-40%) is not dominated by the temporal variations (~5% over a rotation) in the production rate.

As an independent check, the column densities computed for Garradd were compared to the optically thick boundary conditions calculated in a simplified manner for $H_2O$ and $CO_2$ for Tempel 1 by Feaga et al. (2007a) where the effects of two different rotational temperatures (30 K and 100 K) were studied.  The column densities for the central pixel of the HRI-IR Garradd observations listed in Table 3 are of the same magnitude as those in Feaga et al., again indicating that the Garradd analysis should take into account opacity and use the effective g-factor for at least the central pixel.  The column densities in the other Garradd apertures are well below the transition value in Feaga et al. (optical depth = 1) agreeing with the model that in the intermediate and large apertures, the column densities would be underestimated by at most 10% if optically thin conditions were assumed.  CO was not studied in that paper, so a quick independent check on CO could not be done.

**3.3 Production Rates**

Assuming solar-pumped infrared fluorescence, a constant radial outflow with a 1/r distribution, and apertures much smaller than the molecular scale length, total gas production rates were estimated for Garradd using the calculated average column densities from Equation 1 and in Table 3.  The production rate is given by:

$Q = 2rv \; \mathcal{N}_{species}$                                                  Equation 2

where Q is the production rate [molecules $s^{-1}$], r is the effective radius of the aperture [cm], and v is the assumed expansion velocity [cm $s^{-1}$].  The gas velocity is not known for the HRI-IR Garradd observations, so for simplicity a constant expansion velocity of $5 \times 10^4$ cm $s^{-1}$ was assumed.  There is no advantage in improving the degree of accuracy in the velocity as any change in the production rate due to a small systematic uncertainty in the velocity is still well within the stochastic measurement uncertainties that dominate the error bars for these data.  For the snapshot in time covered by the DIF observations, effectively 97 days post-perihelion, production rates were determined for $H_2O$, $CO_2$, and CO (Table 3).  For the large aperture, where opacity was least significant over the aperture and where SNR was best, $Q_{H2O} = 4.6 \pm 0.8 \times 10^{28}$, $Q_{CO2} = 3.9 \pm 0.7 \times 10^{27}$, and $Q_{CO} = 2.9 \pm 0.8 \times 10^{28}$ molecules $s^{-1}$.  If one compares these production rates, which are corrected for optically thick conditions, to those calculated assuming an optically thin coma at

~10,000 km ($4.1 \pm 0.7 \times 10^{28}$, $3.6 \pm 0.7 \times 10^{27}$, and $2.7 \pm 0.7 \times 10^{28}$ molecules s$^{-1}$, respectively) the difference is between 5-12%.

## 4. DISCUSSION

### 4.1 Enriched in CO

As the DIF HRI-IR data provide a single snapshot, comparing production rates over a long baseline, as is done in Figure 9, is necessary to interpret the relative abundances. The compilation of measurements span wavelengths from the UV to the submillimeter, and use multiple methods to derive production rates from parent molecules and from fragment species in varying fields of view. As for many comets, the derived $H_2O$ production rates confirm that Garradd's release of $H_2O$ behaved with a typical rise and fall near perihelion, with less scatter in the post-perihelion data. All of the compiled data are in general agreement and show that the $H_2O$ production peaked ~50 days before perihelion and consistently decreased after that.

The production of CO was expected to behave similar to water, noting parallel volatile release in other comets inside the water snow line (e.g., Hale-Bopp; Biver et al. 1997, Mumma et al. 2003). However, the compilation of CO data overplotted in Figure 9 shows that the CO production rate did not reverse, in fact, the CO production continued to increase through the DIF HRI-IR data, the last known measurement of CO for Garradd during the 2011 apparition. With $H_2O$ decreasing and CO increasing, the CO to $H_2O$ abundance ratio increases dramatically post-perihelion.

The production rates for both $H_2O$ and CO as measured by the DIF HRI-IR, highlighted in Figure 9 in red, are consistent within the uncertainties with other observations acquired near the same post-perihelion time frame. The $H_2O$ production is in line with the decreasing trend defined by others. Moreover, the CO production fits well with the surprising monotonic increase seen through perihelion from ground based observations. Together, the continued decrease in water and increase in CO measured by the HRI-IR result in the highest observed abundance ratio occurring 97 days post-perihelion. The absolute production rate ratio of CO to $H_2O$ calculated for Garradd at 2 AU post-perihelion, ~60% ± 20%, is the highest inside 3 AU ever measured for any comet. Previous CO-rich comets range from 10-30% including C/1975 V1-A West (Feldman & Brune 1976), C/1995 O1 Hale-Bopp (Biver et al. 1997, DiSanti et al. 1999), C/1999 T1 McNaught-Hartley (Mumma et al. 2003), C/1996 B2 Hyakutake (Mumma et al. 1996, Biver et al. 1999, DiSanti et al. 2003), and C/2008 Q3 Garradd (Ootsubo et al. 2012).

### 4.1.1 Garradd's Water Release

The $H_2O$ production rates derived from the HRI-IR data are compared to many other measurements in Figure 9, including the OH narrow band data acquired by the DIF MRI camera. The MRI data were taken slightly before the IR data were acquired and the Vectorial Model (Festou 1981), with an assumed parent outflow velocity of 0.588 km s$^{-1}$, was used to estimate $H_2O$ production rates from the integrated OH brightness. The absolute $H_2O$ production rate for the large HRI-IR aperture is 30% smaller than the MRI $H_2O$ production rate deduced from the measured surface brightness of OH in an aperture with radius of $1.3 \times 10^5$ km (Farnham et al. 2013, in preparation). This discrepancy can be explained by a combination of the following: the MRI OH filter data acquired closest in time with the IR data were collected 24 days prior to the IR data and the production rate trend for water in other observations indicates a decrease in water of ~30% during that time; the radius of the MRI field of view was an order of magnitude larger and encompassed the entire detectable signal in the MRI; and the assumed parent outflow velocity was 15% larger than that used in the IR derivations. Therefore, well within the error bars, the DIF MRI and HRI-IR data are consistent. The MRI data points are included in Figure 9.

Other longer baseline measurements and $H_2O$ production rates are also in agreement with each other and the DIF data. $H_2O$ production rates deduced from broadband Swift UVOT measurements of OH span from nearly 250 days pre-perihelion through 150 days post-perihelion (Bodewits et al. 2013, in preparation). $H_2O$ production rates from the Swift measurements, acquired just a few days after the HRI-IR data, were derived for a 55,000 km radius aperture (5 times larger than HRI-IR) using the Vectorial Model (Festou 1981), again with an assumed parent outflow velocity of 0.588 km s$^{-1}$. The Swift data are in good agreement with data from all other observers, including the HRI-IR data, and show that the $H_2O$ production continues to drop beyond the timeframe of the DIF HRI-IR observations (Figure 9).

Combi et al. (2013) presented a wealth of data covering Garradd's apparition for ±100 days. Their production rates were derived from data acquired by SOHO's SWAN Lyman-α camera and are systematically larger than all other observers. The field of view of that instrument is large, 5° by 5°, and the aperture size used in the analyses is $2 \times 10^7$ km, three orders of magnitude larger than the largest HRI-IR aperture. The SWAN $H_2O$ production rates are a factor of two higher than the HRI-IR, MRI and other contemporaneous data (Figure 9). A difference in aperture sizes could explain the systematic discrepancy between SWAN and all other observers, with the large SWAN apertures capturing all of the hydrogen in the comet, independent of the hydrogen's parent molecule or initial state. Combi et al. (2013) suggest that icy grains, having had time to sublimate in the large SWAN field of view, could account for much of the discrepancy with other observers, especially pre-perihelion. Combi et al. also exclude the contribution of other common cometary hydrogen bearing species to the hydrogen abundance, which could account for a significant inflation of the hydrogen population. Combi et al. determined that the post-perihelion decrease in H$_2$O production was fit by a power law with a slope of -3.2, following the expected variation for water sublimating from a constant surface area, i.e. the nucleus and not icy grains.

Additional narrow band OH measurements made at Lowell Observatory are presented in Farnham et al. (2013, in preparation) and shown in Figure 9. The observations spanned from 220 days pre- to 120 days post-perihelion, and give H$_2$O production rates that are consistent with many other observers, including the HRI-IR values acquired within a month before and after two nights of observing at Lowell. The radius of the aperture used in the analysis varied between 25,000 km and 135,000 km and the parent velocity for the H$_2$O molecules was taken to be ~0.6 km s$^{-1}$ for the data nearest in time with the HRI-IR data.

The other H$_2$O production rates in Figure 9 were assembled from Villanueva et al. (2012b), DiSanti et al. (2013), Paganini et al. (2012), Bockelée-Morvan et al. (2012), and Feldman et al. (2012). Villanueva et al. (2012b) presented direct infrared detections of H$_2$O from Keck II NIRSPEC pre-perihelion. To derive production rates, they assumed an outflow velocity of 0.55 km s$^{-1}$ in a 1,900 km by 455 km field of view. DiSanti et al. (2013) also discussed detections of H$_2$O (pre- and post-perihelion) acquired with Keck II NIRSPEC. Their production rates were computed using a growth factor formalism with the molecular distribution along the narrow slit out to 4,000 km and assuming outflow velocities of 0.59 km s$^{-1}$ and 0.65 km s$^{-1}$, pre- and post-perihelion respectively. Paganini et al. (2012) reported pre-perihelion direct measurements of H$_2$O from the CRIRES instrument at VLT within a 9,000 km by 450 km field of view. They assumed a constant outflow velocity of 0.6 km s$^{-1}$ in order to calculate production rates. Bockelée-Morvan et al. (2012) presented pre-perihelion H$_2$O production rates calculated directly from Herschel HIFI data out to 100,000 km from the nucleus with a constant expansion velocity of 0.6 km s$^{-1}$. Feldman et al. (2012) estimated post-perihelion H$_2$O production rates from HST STIS observations of OH in the central 250 km by 620 km of the slit with an assumed parent velocity of 0.7 km s$^{-1}$.

**4.1.2 Garradd's Carbon Monoxide Release**
There are many fewer CO observations for Garradd than the copious H$_2$O measurements. Even so, direct detections of CO and derived production rates for Garradd are available during the time period spanning 150 days pre-perihelion through 100 days post-perihelion and are plotted in Figure 9. Unfortunately, there are no CO observations within ± 30 days of the HRI-IR data. Nonetheless, a clear trend of monotonically increasing CO production throughout the time period is evident. Such behavior has never before been observed for any volatile in any comet near perihelion. The HRI-IR post-perihelion CO production rate, corrected for opacity, at 97 days is consistent with this trend and is compared to the other CO measurements below and in Figure 9.

Yang & Drahus (2012) at the 15-m James Clerk Maxwell Telescope (JCMT) conducted the only study with a long temporal baseline of Garradd's CO production.

The data were acquired with the 0.8 mm HARP receiver in a position-switching mode. CO was observed during four runs: 2011 Jul 28-31 (r=2.48 AU), 2011 Sep 23-25 (r=1.97 AU), 2011 Oct 23-25 (r=1.76 AU), and 2012 Jan 6-8 (r=1.56 AU). The first three observing runs were pre-perihelion and the last run was post-perihelion. Isotropic production rates determined from the velocity-resolved spectra of the optically thin J=3-2 transition are presented in Table 4 and plotted in Figure 9. The production rates were obtained simultaneously with the expansion velocities by fitting the optically thin line shape model described by Drahus et al. (2010, 2012). The model accounts for finite molecular lifetime and assumes radial outflow at a constant velocity and LTE at a constant temperature. The temperature is either determined from observing multiple methanol lines at JCMT or adopted from literature. The least-squares fit shows that isotropic outgassing models the data much better than the asymmetric model in which outgassing comes from the day-side only. The uncertainties were estimated using a Monte Carlo method. The error in the production rate also accounts for the poorly characterized loss of the telescope's main beam efficiency due to differential heating of the antenna during daytime observations. Moreover, the JCMT monitoring of the bright HCN line in Garradd did not show any obvious short-term variation, such as that observed in comet 103P/Hartley 2 (Drahus et al. 2011, 2012), hence the rotational modulation of CO was most likely negligible within the precision. For full details of the JCMT observations of comet Garradd see Yang & Drahus (2013, in preparation).

Individual CO production rates displayed in Figure 9 were assembled from a variety of sources. Villanueva et al. (2012b) presented NASA IRTF CSHELL infrared detections of CO and calculated pre-perihelion production rates for Garradd in a 4,450 km by 2,120 km field of view assuming a constant outflow velocity of 0.55 km s$^{-1}$. Paganini et al. (2012) reported pre-perihelion measurements of CO from CRIRES, simultaneous to their $H_2O$ detections, within a 9,000 km by 450 km field of view and estimated production rates assuming a velocity of 0.6 km s$^{-1}$. DiSanti et al. (2013) discussed a pre-perihelion detection of CO acquired with Keck II NIRSPEC and computed production rates using a growth factor formalism with the molecular distribution along the narrow slit out to 4,000 km and assuming outflow velocities of 0.59 km s$^{-1}$. Feldman et al. (2012) estimated post-perihelion CO production rates from HST COS observations of the CO Fourth Positive system in a 2,880 km diameter aperture assuming a velocity of 0.7 km s$^{-1}$ and taking into account saturation of individual lines following the approach of Lupu et al. (2007) for UV bands. Biver et al. (2012) observed CO in Garradd post-perihelion with the IRAM 30 m radio telescope, derived an expansion velocity of 0.65 km s$^{-1}$, and calculated a post-perihelion production rate from the data. All of these observations are consistent with a continuing increase of CO through the HRI-IR observations.

**4.2 Carbon Dioxide Production Rate**

The DIF HRI-IR production rate for $CO_2$, $3.9 \pm 0.7 \times 10^{27}$ molecules s$^{-1}$, is the only direct measurement of $CO_2$ for comet Garradd. This production rate results in a typical cometary coma abundance ratio of 8% as compared to $H_2O$. With only one measurement for $CO_2$ in Garradd, no firm trend can be established for its release during the 2011 apparition to compare with the $H_2O$ or CO outgassing behavior. Decock et al. (2013) measured the forbidden oxygen lines in comet Garradd from 3.25 AU to 2.07 AU pre-perihelion and reported a systematic decrease in the green to red line ratio. These results suggest that the $CO_2$ does not trend with the water as it increases to its maximum activity, perhaps indicating that $CO_2$ activity follows the CO trend instead or remains relatively constant through the apparition.

**4.3 Volatile Behavior of Dynamically Young Comets**

In Hale-Bopp, the production rates of all volatiles were seen to vary nearly in parallel (Biver et al. 1997, Mumma et al. 2003) until the comet was beyond the snow line of $H_2O$. Outside that point, hypervolatiles drove the activity and their production rates dominated over water but $H_2O$ and CO continued to decrease. Unlike Hale-Bopp, the release of CO from Garradd monotonically increased by over an order of magnitude from heliocentric distances of 2.5 AU pre-perihelion to 2 AU post-perihelion, while the $H_2O$ peaked pre-perihelion and then monotonically decreased. Garradd's asymmetric activity suggests a seasonal effect with two separate active regions or volatile reservoirs with differing composition on different hemispheres, one rich in $H_2O$, the other enriched in CO, being activated at different times based on the orientation of the rotational pole of the nucleus to the Sun and the seasonal protection from the thermal wave propagation through the nucleus. Or, possibly this outgassing behavior is a property of relatively dynamically young comets like Garradd where a rapid pre-perihelion mass loss of $H_2O$ may have unveiled deeper, less altered material through the perihelion passage where more volatile species could have survived. The composition revealed post-perihelion may be closer to the original bulk composition of formation. Perhaps a 60-75% abundance ratio of CO to $H_2O$ is a more typical composition of comets at formation or perhaps it is indicative that Garradd was originally formed farther beyond the CO snowline than most other comets.

The CO abundance in the two most dynamically young comets in the A'Hearn et al. (2012) study, C/1989 X1 Austin and C/2007 N3 Lulin, were only observed a few times and at small heliocentric distances. However, they showed a low CO to $H_2O$ ratio. This is counterintuitive since younger comets should be less depleted in their volatiles than comets that have come close to the Sun multiple times. If other dynamically new comets exhibit behavior similar to Garradd, it is possible that the $H_2O$ was at or near peak production during the observations of Austin and Lulin but the CO production had not yet peaked. Without a set of longer baseline observations of dynamically young comets Austin and Lulin, and other dynamically new comets, we can only hypothesize that hypervolatiles at depths in dynamically young comets may be preferentially released later in time than $H_2O$ and with an increase in activity as the thermal wave penetrates the depths of the reservoirs well beyond perihelion passage. Another possibility is that these comets are in fact different in composition from Garradd and formed over a wide range of heliocentric distances in the proto-planetary disk, causing the primordial CO abundance to be divergent.

### 4.4 Asymmetries Along the Garradd-Sun Line

It is expected that volatiles, especially $H_2O$, sublimate preferentially in the sunward direction when warmed by the Sun, but nightside jets and activity driven by $CO_2$ on Tempel 1 and Hartley 2 (Feaga et al. 2007a, Farnham et al. 2007, A'Hearn et al. 2011) show that is not always the case. Villanueva et al. (2012b) and Paganini et al. (2012) presented sunward enhancements for Garradd's pre-perihelion $H_2O$ and OH profiles, but in contrast, identified an anti-sunward enhancement for CO, HCN and $CH_4$.

The coma distribution was investigated in the HRI-IR observations along the Sun-anti-Sun line, i.e. in the direction along the slit, in order to explore asymmetries that may exist in the data. A sunward spectrum and an anti-sunward spectrum were extracted and are presented in Figure 10. Each spectrum consists of 4 pixels along the slit excluding the central pixel, extending 8,420 km sunward and 8,420 km in the anti-sunward direction, respectively. This distance corresponds to the maximum extent along the slit in the large aperture data. The spectra show more intense $H_2O$ and $CO_2$ bands in the sunward direction as compared to the anti-sunward direction although the enhancements are not statistically significant (Table 5). At the wavelengths of CO, the spectrum extracted from the sunward direction is dominated by noise whereas in the anti-sunward spectrum, there is a single peak above the noise. However, when integrated over the CO band wavelength bin, the average surface brightness of the noise in the sunward direction is larger than the average anti-sunward surface brightness in the peak and the large uncertainties overwhelm the values (Table 5). Thus, the HRI-IR data suggest, but do not convincingly show, a pronounced sunward vs. anti-sunward asymmetry, with $H_2O$ and $CO_2$ enhanced in the sunward direction and the CO band more well defined in the anti-sunward direction similar to the findings of Villanueva et al. (2012b) and Paganini et al. (2012).

### 4.5 Unidentified Feature at 2.8 µm

A strong emission feature was detected in the data between 2.75 and 2.85 µm, peaking at 2.8 µm, longward of the water, and has not been seen in other DIF HRI-IR cometary flyby data. The feature has a distinct spatial profile that is different from both the $H_2O$ and $CO_2$ distribution. It is enhanced in the anti-sunward direction, the bulk of the intensity is along the scan direction in the intermediate annulus, and the feature is not present above the noise in the scans acquired on the first day of observations. The possible spectral sources listed below were identified but rejected for reasons given. If the feature were due to hot water bands or OH prompt emission, the distribution should map more closely with the distribution of $H_2O$, with increased intensity where the $H_2O$ column density is largest. In addition, the integrated flux associated with the feature is an order of magnitude too large to be OH fluorescence and hot water bands and OH should be broader than the detected feature with emission between 2.85 and 3.0 μm (Villanueva et al. 2012a, Schleicher & A'Hearn 1988). It is plausible that the emission feature is due to an ion, which would be more consistent with the tailward spatial distribution. $H_3O^+$ has been detected with mass spectrometers at comet 1P/Halley and 19P/Borrelly (Krankowsky et al. 1986, Nordholt et al. 2003) and its $\nu_3$ ro-vibrational band is centered at 2.83 μm. However, the strong feature is only present in scans acquired on the second day of observing and since the observing geometry and strength of the $H_2O$ emission did not change between the two days, an ion tail explanation is not likely.

It is more plausible that the feature is an artifact, so the data were examined more closely in smaller temporal chunks to search for the introduction of an artifact into the data. The tailward distribution was very prominent in the data in the last eight hours of the scans acquired on the second day of observations. Because similar phase coverage and a full rotation of the comet was sampled during each of the two days of observation, it is unlikely that the feature is due to a rotational effect or it would have been seen on both dates. No obvious bad pixel was found in the frames and the location of the comet moved by 15 pixels along the slit during the set of scans. As such, a single bad pixel or group of pixels should not dominate the spatially co-aligned and resistant-meaned data. No cosmic ray hits were noted in the scans, but it is possible that a damaging cosmic ray hit the detector between scans on 2012 Apr 2 and created an artifact at 2.8 μm. Lastly, stable excess noise lasting for several hours and caused by defects in the vicinity of p-n junctions in HgCdTe detector HAWAII-1R multiplexers, like the sensor on the HRI-IR, is well documented (Bacon et al. 2005). This noise can manifest as a region of neighboring pixels measuring a steady signal and is the most likely cause of the 2.8 μm feature. Combined, these lines of reasoning suggest that the relatively strong 2.8 μm emission is due to an artifact.

## 5. CONCLUSIONS

The Deep Impact Flyby spacecraft's HRI-IR spectrometer detected $H_2O$, $CO_2$, and CO in the coma of comet C/2009 P1 (Garradd). The simultaneity of the detections reduced systematic effects when comparing absolute abundances. Garradd was observed with the HRI-IR 97 days post-perihelion when it had reached a heliocentric distance of 2 AU. The most accurate production rates were derived with modeled g-factors for the large aperture having the smallest optical depth effects. The opacity corrected $H_2O$ production rate, $Q_{H2O} = 4.6 \pm 0.8 \times 10^{28}$ molecules s$^{-1}$, is consistent with contemporaneous measurements while that of CO, $Q_{CO} = 2.9 \pm 0.8 \times 10^{28}$ molecules s$^{-1}$, is consistent with the trend seen in the CO measurements made by other observers. The $CO_2$ production rate, $Q_{CO2} = 3.9 \pm 0.7 \times 10^{27}$ molecules s$^{-1}$, and thus abundance ratio of 8% to $H_2O$, is typical for comets and is the only directly measured $CO_2$ value for Garradd. The HRI-IR data are consistent with an optically thick coma transitioning to optically thin around 10,000 km. The HRI-IR data also validate that a 15-20% increase in gas velocity in the inner coma as Combi (2002) modeled for Hale-Bopp, is plausible for Garradd. The HRI-IR data suggest a similar sunward/anti-sunward asymmetry in the behavior of the $H_2O$ vs. CO outgassing as Villanueva et al. (2012b) and Paganini et al. (2012) found, but are not conclusive.

The HRI-IR direct detection of $H_2O$ and the derived production rate verify the post-perihelion decrease in outgassing seen by other observers while the direct detection of CO and the retrieved production rate solidify a continuing increase in CO outgassing through perihelion. The behavior of Garradd's $H_2O$ outgassing, increasing and peaking pre-perihelion and then steadily decreasing, is more typical than the unusual behavior of CO, which monotonically increased throughout the entire apparition. This uncorrelated volatile behavior around perihelion could be due to a seasonal effect where the two hemispheres of Garradd were heated at different rates with one hemisphere heated by the Sun pre-perihelion and the other thermally protected. Post-perihelion, the hemispheres, composed of cometesimals formed at different distances from the Sun such that one is more enriched in CO than the other, could have rotated and the CO enriched hemisphere became active as the thermal wave propagated through. Or, possibly this outgassing behavior is a property of relatively dynamically young comets like Garradd where a rapid pre-perihelion mass loss of $H_2O$ may have unveiled deeper, less altered material through the perihelion passage where more volatile species could have survived and then began to sublime with an increasing rate. Whatever the cause for Garradd's unusual volatile release, when measured by the HRI-IR 97 days post-perihelion, Garradd exhibited the highest CO to $H_2O$ abundance ratio ever observed for any comet inside the water snow line at ~60%. This study explicitly shows how critical it is to have long baseline observations of comets to examine their volatile release if we expect to understand true primordial compositions.

**ACKNOWLEDGEMENTS**

This work was funded by NASA, through the Discovery Program, via contract NNM07AA99 C to the University of Maryland and task order NMO711002 to the Jet Propulsion Laboratory. The Jet Propulsion Laboratory is operated by the California Institute of Technology. The authors want to thank all of the Garradd observers with whom we've had insightful discussions over the past two years at various conferences and meetings. B. Yang was supported through the NASA Astrobiology Institute under Cooperative Agreement No. NNA08DA77A issued through the Office of Space Science. M. Drahus is a Jansky Fellow of the National Radio Astronomy Observatory, a facility of the National Science Foundation operated under cooperative agreement by Associated Universities, Inc. D. Schleicher gratefully acknowledges support by NASA's Planetary Astronomy Program.

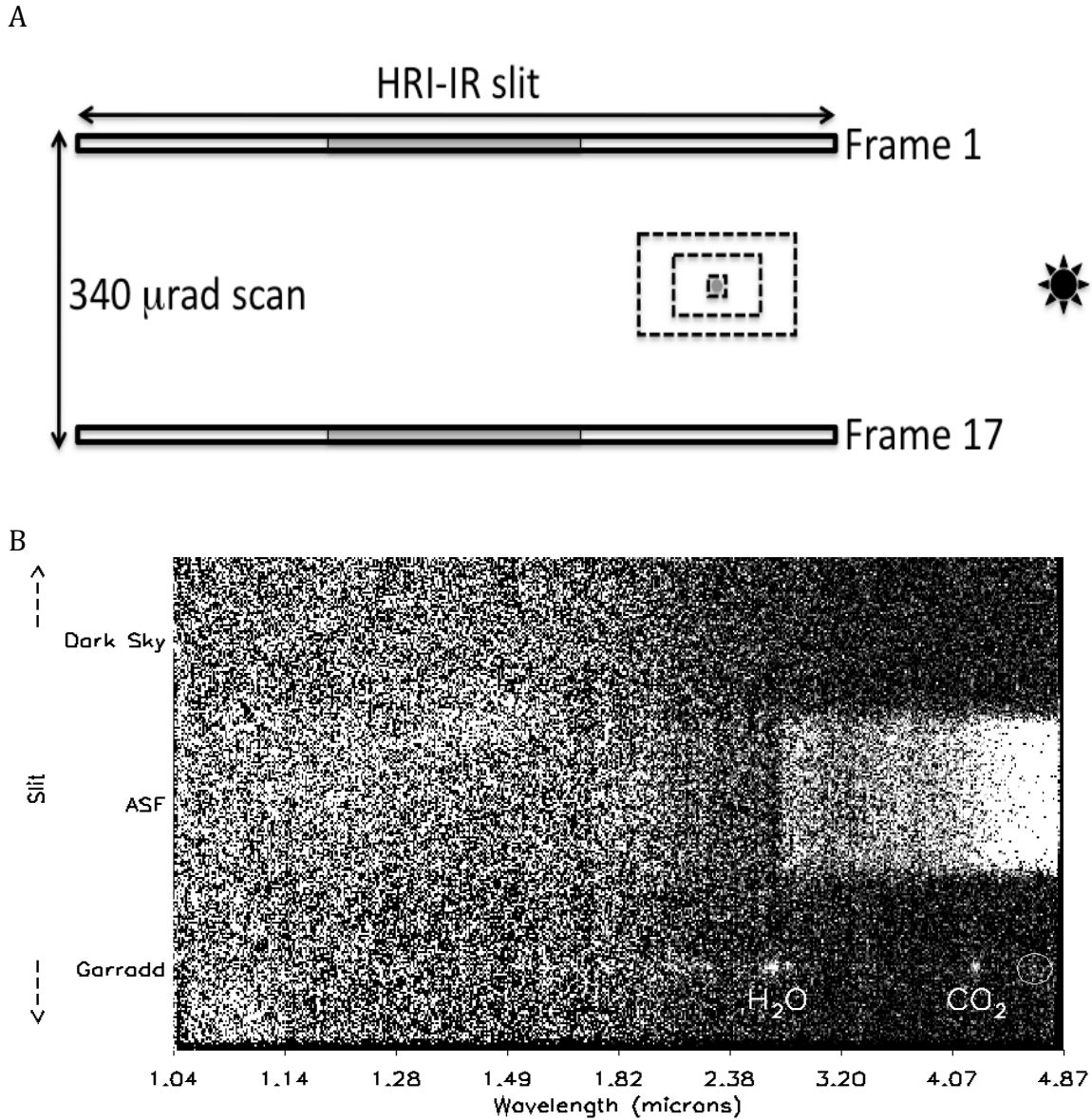

Figure 1 – (A) Scan orientation diagram detailing the slit orientation with respect to the projected comet-Sun line as well as the scan direction and spatial extent of a 17-frame scan. The aspect is rotated 90º relative to that of Panel (B) for presentation purposes. For each scan, Garradd was targeted to fall in the third of the detector closest to the Sun in frame 9. Due to pointing uncertainty, the comet was not always in frame 9. The diagram is not drawn to scale, as the slit is 2560 μrad in length while the scan covers only 340 μrad. The dashed boxes show the size (exaggerated for display purposes) and placement of the rectangular apertures and annuli that are analyzed in the paper. The Sun is to the right in the diagram. (B) Representative spatial-spectral surface brightness distribution map of comet Garradd created by averaging 5 frames in the scan direction. The spectral dimension is horizontal and the spatial (along the slit) direction is vertical in this diagram. Garradd is located in a row near the middle of the bottom third of the detector. The spatial extent of Garradd's signal is about 5 pixels radially from the central pixel. The $H_2O$ (2.67 µm) and $CO_2$ (4.26 µm) bands are clearly seen with a suggestion of the CO (4.67 µm) band near the edge of the detector and a likely instrumental artifact just longward of the $H_2O$. The stretch is linear from 0 to $2 \times 10^{-5}$ W m$^{-2}$ sr$^{-1}$ µm$^{-1}$. There is a lot of noise across the field of view, especially on the short wavelength side of the detector. Below ~2 µm, the detector has very low sensitivity and low DN from noise get magnified in the conversion to radiance. The central third of the slit shows an artifact due to the anti-saturation filter in the instrument meant to prevent saturation of the detector when imaging a resolved nucleus. Although optimally removed in higher SNR data, this artifact is very prominent in very low signal data and especially at the long wavelength side of the detector where the throughput of the filter drops to zero (Klaasen et al. 2013). Garradd was intentionally positioned outside of the anti-saturation filter to avoid this region.

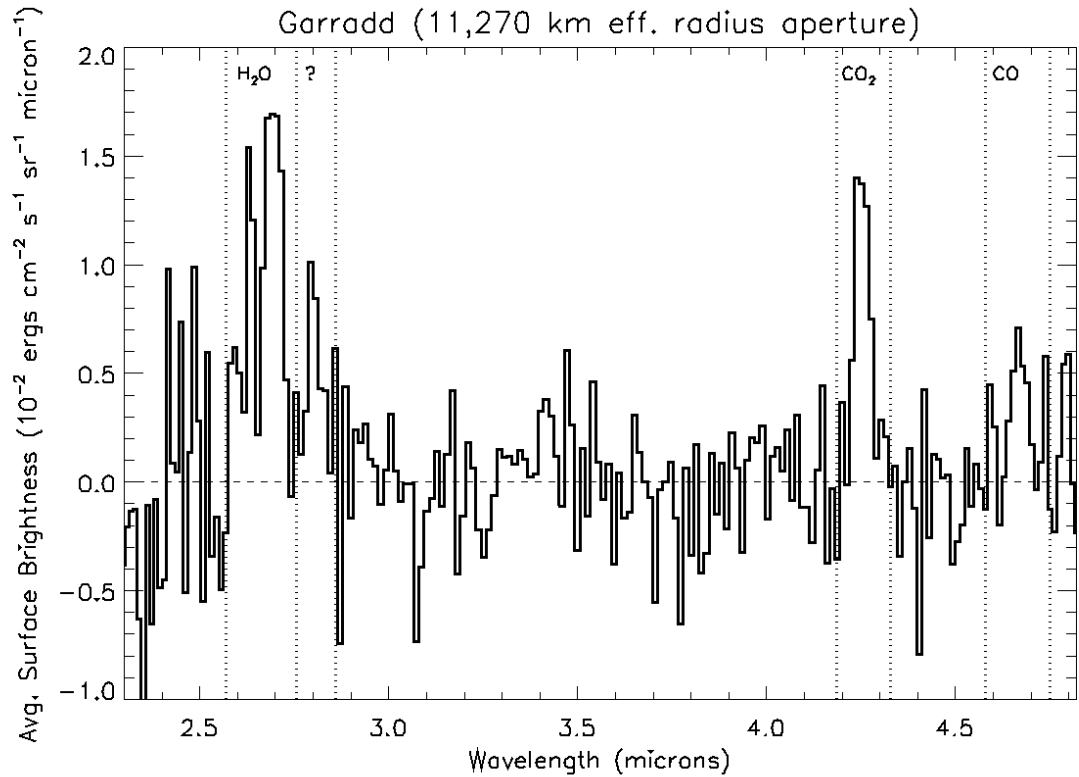

Figure 2 – Spectrum of comet Garradd extracted from the large aperture with an area of 21,050 km by 18,945 km projected at the comet. The unresolved nucleus is contained in the central pixel of the box aperture. Emission features are present for the dominant parent volatiles, $H_2O$, $CO_2$, and CO. Vertical dotted lines indicate the bounds of integration for the total flux of the emission bands. The "?" indicates an unidentified feature, most likely an instrumental artifact.

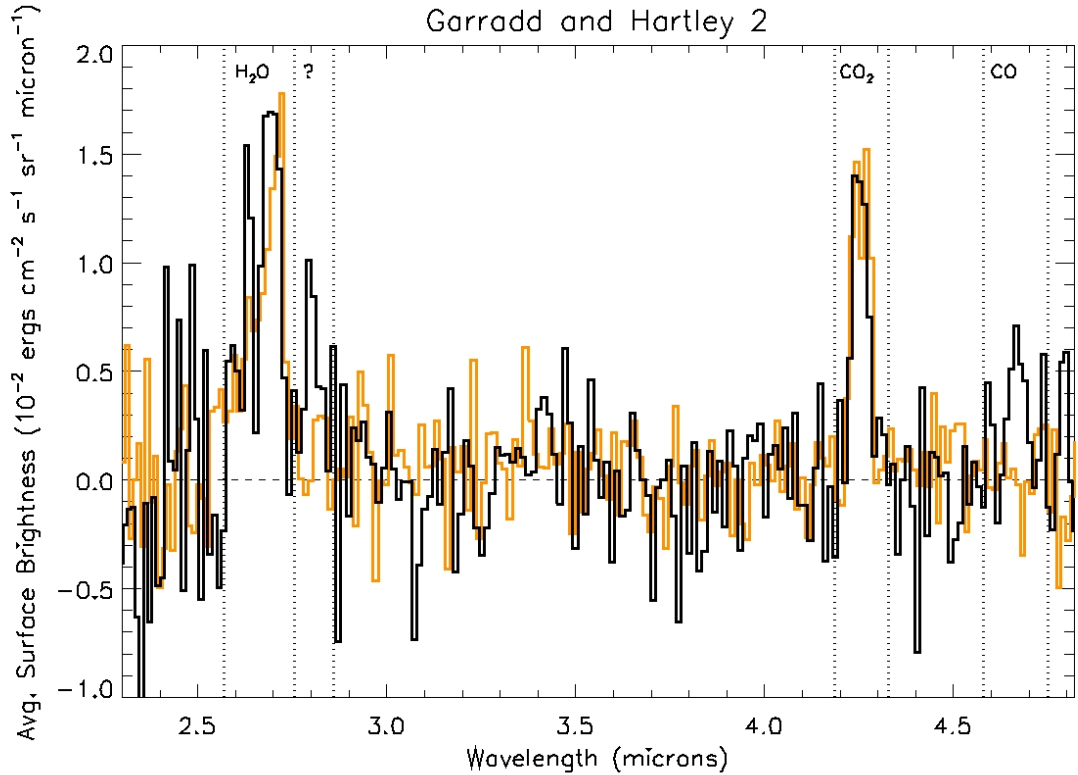

Figure 3 – Spectrum of comet Garradd (black) overplotted with a representative spectrum of comet 103P/Hartley2 (orange, from 2010 Oct 16). Both observations were acquired with the DIF HRI-IR spectrometer, using the same observing technique, and have been processed with the same pipeline, so the comparison should have minimal systematic errors. Comet Hartley 2 is known from independent data to be depleted in CO at heliocentric distances near perihelion (Weaver et al. 2011). Garradd, on the other hand, is enriched in CO. The CO emission band can be clearly seen in the Garradd spectrum (black) as compared to the Hartley 2 spectrum (orange). Similarly, an unidentified emission band, most likely an artifact, is apparent in Garradd but not in Hartley 2 at 2.8 μm. The spectrum of Hartley 2 was reduced by a factor of 7.8 to match the peak $CO_2$ surface brightness of Garradd.

A

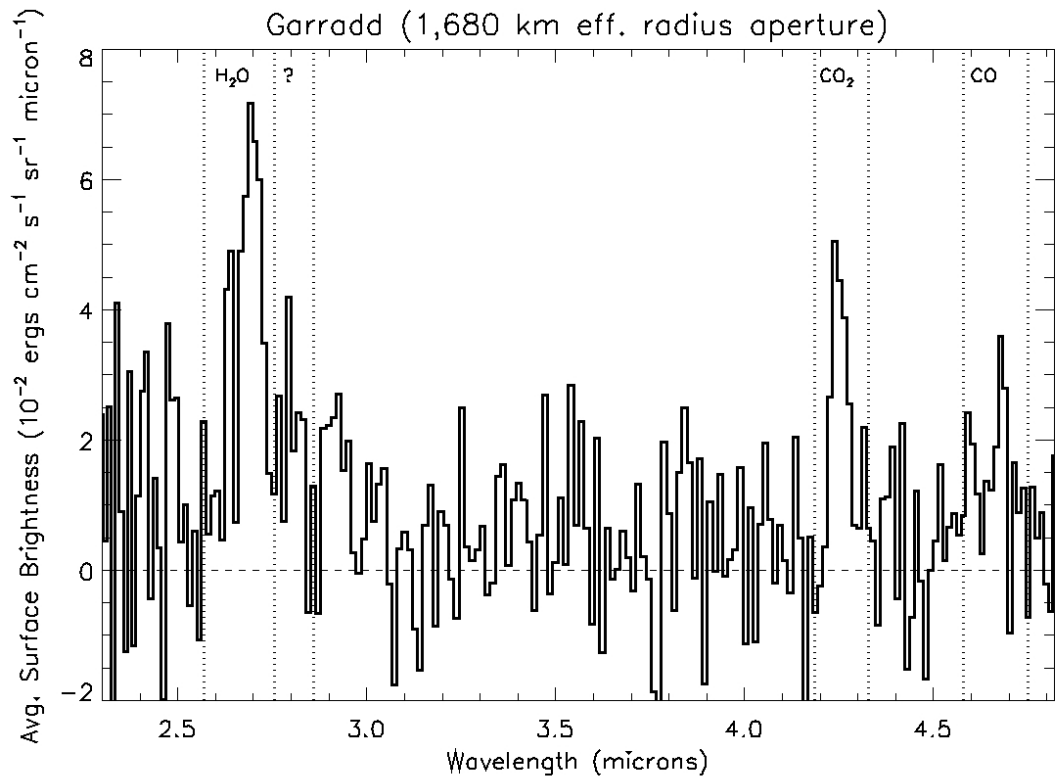

B

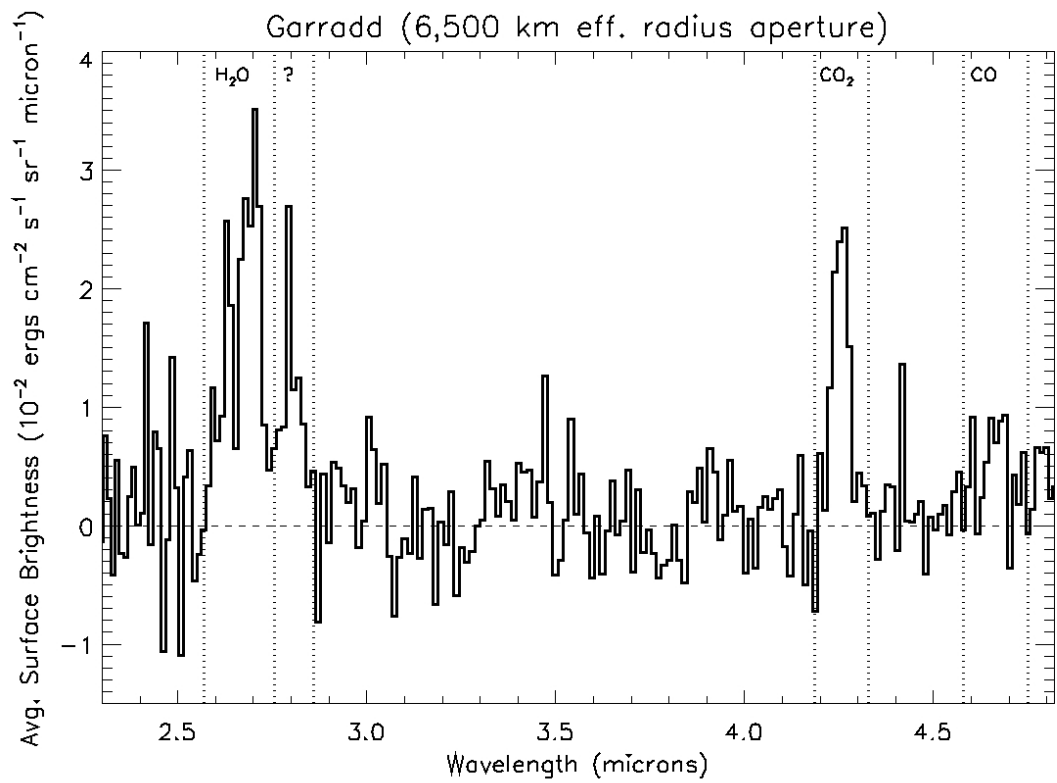

Figure 4 – Same as Figure 2 for (A) the small aperture (4,210 km by 2,105 km area) and (B) an intermediate aperture (12,630 km by 10,525 km area). The pixel containing the nucleus is in the central pixel of each of these boxes.

A

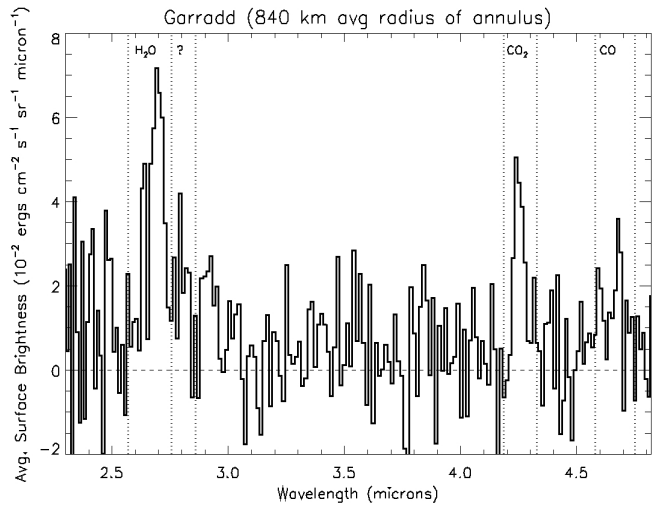

B

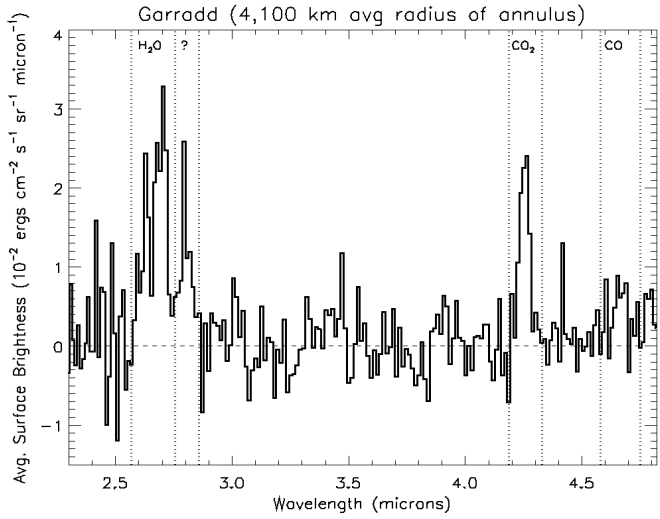

C

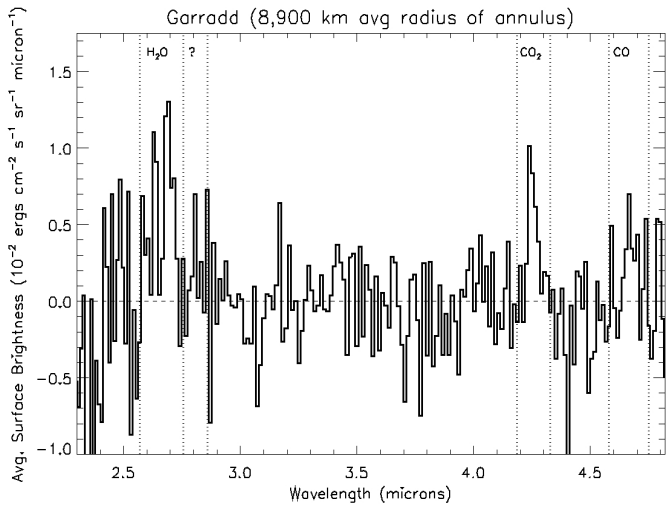

Figure 5 – Spectra extracted for comet Garradd in three different rectangular annuli, corresponding to average annular radii of about 840 km (A), 4,100 km (B), and 8,900 km (C).

A

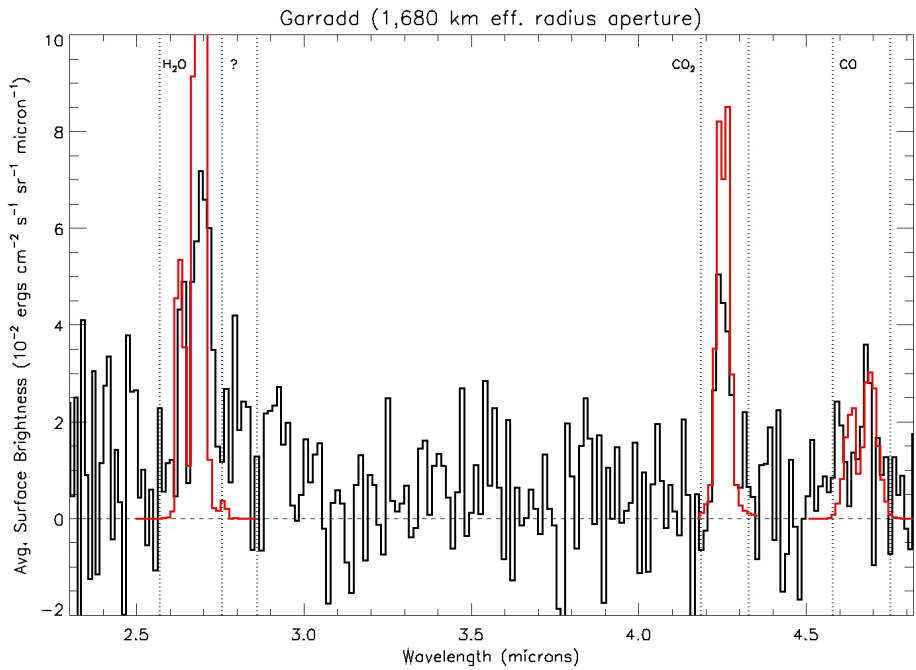

B

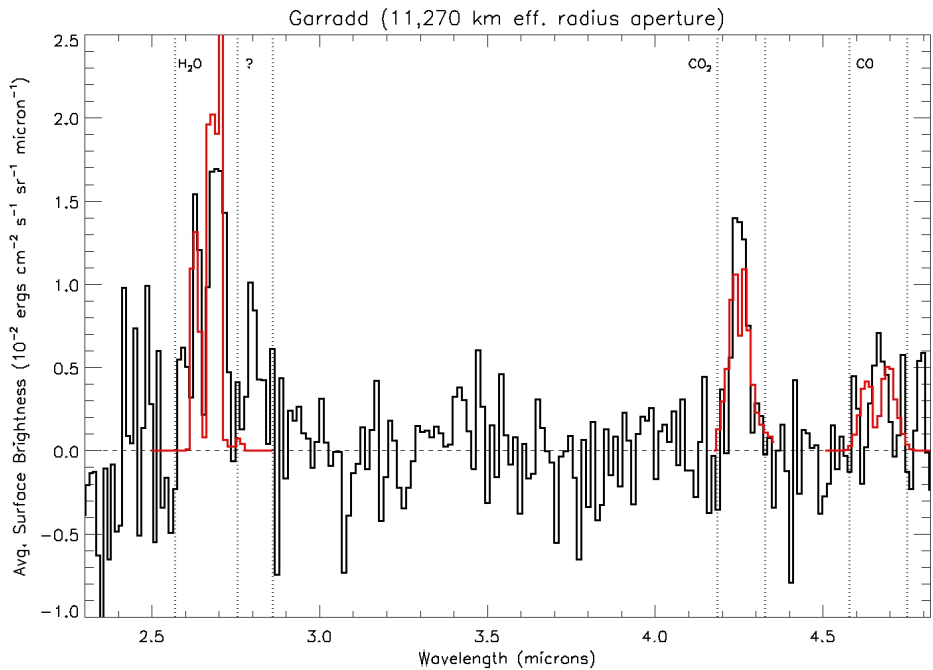

Figure 6 – Small (A) and large (B) aperture spectra presented in Figures 4 and 2, respectively, with modeled surface brightness overplotted to show the fit and appropriateness of the modeled optically thick g-factors. Integrated band fluxes of the model and the data match to within 3% for the largest apertures where the discrepancy in band peaks between the model and data are within the noise of the data.

A

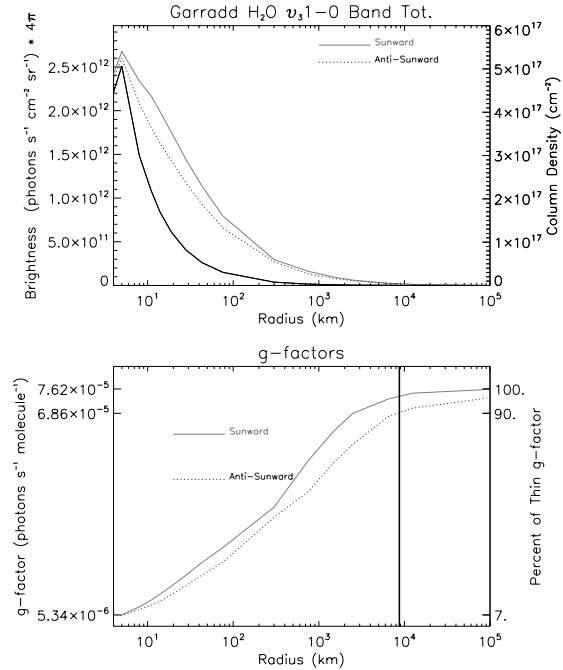

B

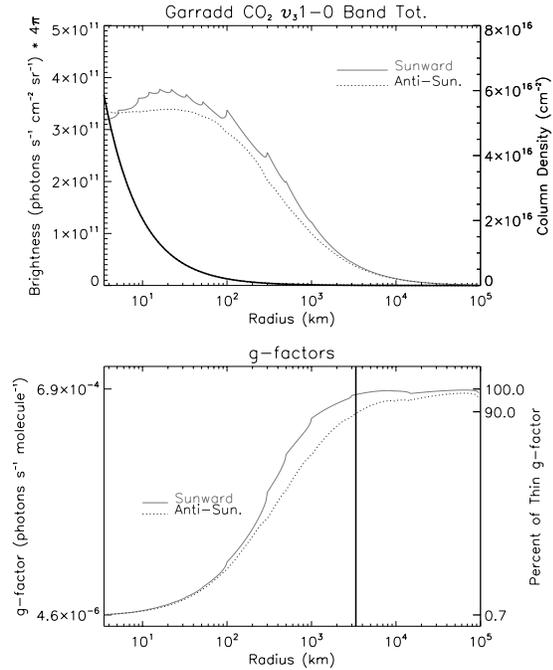

C

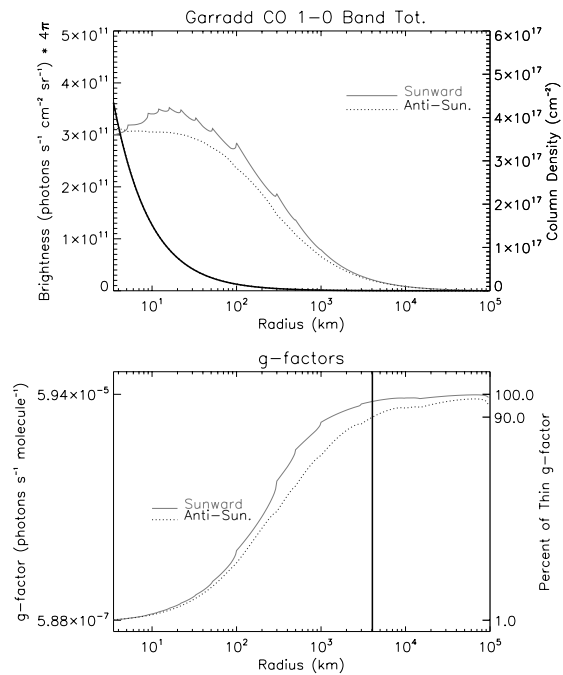

Figure 7 – The spherically symmetric radiative transfer model for Garradd predicts the variation of brightness, column density (top plot for each molecular species) and effective g-factor (bottom plot for each species) with distance allowing for effects of optical depth. The solid black curve in each of the brightness plots shows how the column density of each species varies with distance from the nucleus while the sunward (solid gray) and anti-sunward (dotted) curves show the modeled brightness. The coma transitions from optically thick to thin ($g > 0.9\ g_{thin}$) beyond 9,000 km, 3,000 km, and 4,000 km from the nucleus for (A) $H_2O$, (B) $CO_2$ and (C) CO, respectively, where the g-factors are at least 90% complete as shown by the solid vertical line. g-factors along a line of sight on the sunward side of the coma and one on the anti-sunward side are shown approaching the asymptotic g-factor calculated in the model.

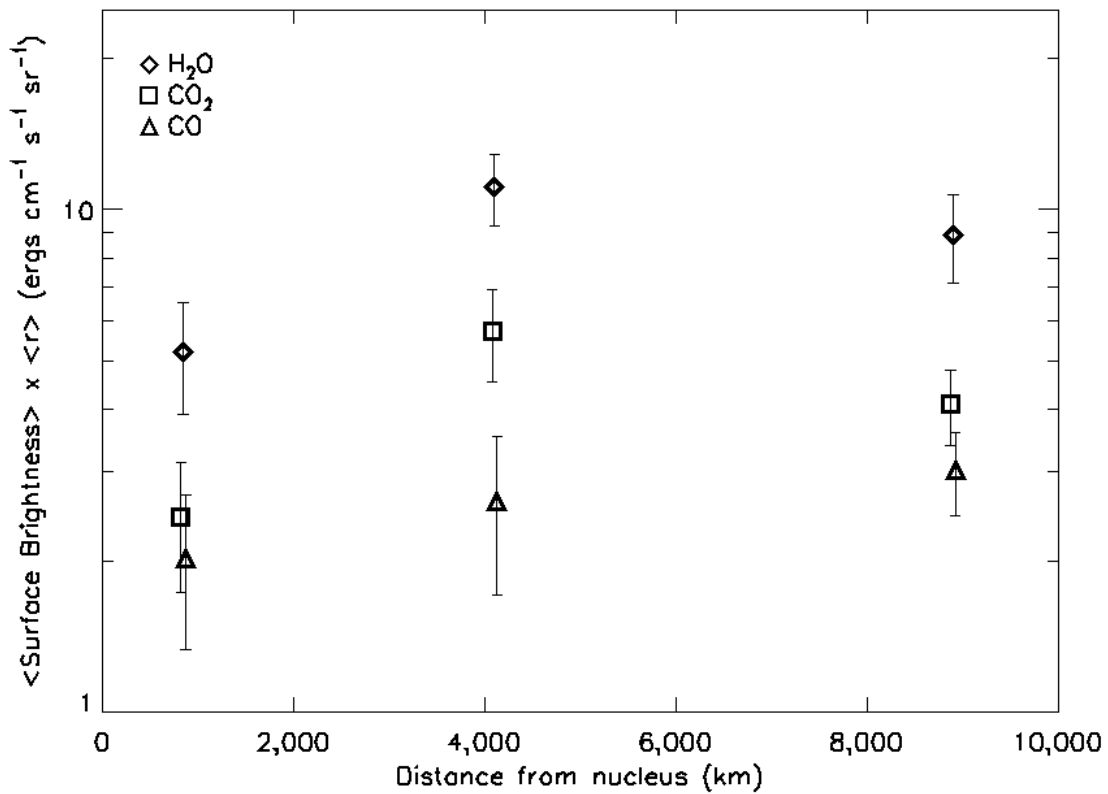

Figure 8 - Plot of average integrated surface brightnesses multiplied by the average size of the rectangular annuli to show the effects of opacity on the signal detected from various annuli. If all three annuli were optically thin, the data should be constant with distance and a flat line could be drawn through them. For CO, while the central pixel value is less than that for the other annuli, all three values are within the measurement uncertainties. However, the first measurement, i.e. the central pixel, is about a factor of two less than the measurements in the other annuli for $H_2O$ and $CO_2$. This underestimate of the average surface brightness is a result of optical depth effects. The deviation from a constant product of surface brightness and size of the annulus with distance between the medium and large annuli could be attributed to relative optical depth effects between the three molecules. The deviation is also consistent with erroneously assuming a constant outflow velocity in the inner coma as has been done here for Garradd. The deviation is too large to be caused by temporal changes in production rates alone.

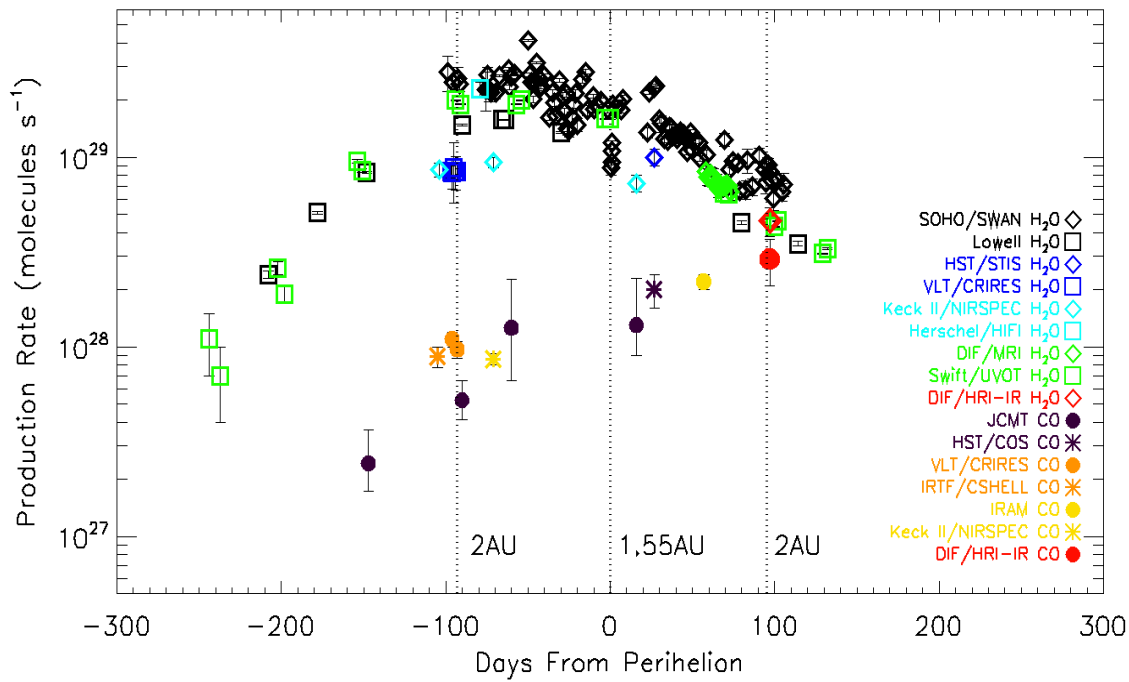

Figure 9 – Plot showing the trend in production rates for $H_2O$ (diamonds and squares) and CO (filled circles and asterisks) around comet Garradd's 2011 perihelion. These data points were collected from many observers and many observatories. The DIF HRI-IR production rates computed for $H_2O$ and CO in this paper are highlighted in red. SOHO SWAN Ly-a measurements were reported by Combi et al. (2013), Lowell narrow band OH observations obtained by Schleicher (Farnham et al. 2013, in preparation), HST COS and STIS UV detections of OH and CO by Feldman et al. (2012), ESO VLT CRIRES observations of $H_2O$ and CO by Paganini et al. (2012), Keck II NIRSPEC and IRTF CSHELL $H_2O$ and CO data by Villanueva et al. (2012b), Keck II NIRSPEC $H_2O$ and CO data by DiSanti et al. (2013), Herschel HIFI detections of $H_2O$ by Bockelée-Morvan et al. (2012), Swift UVOT observations of OH by Bodewits (2013, in preparation), DIF MRI narrow band OH measurements by Farnham et al. (2013, in preparation), JCMT submillimeter detections of CO by Yang & Drahus (2012, updated results presented in this paper), and IRAM 30-m CO data by Biver et al. (2012 and private communication). The water production rate peaks around 50 days pre-perihelion while the CO appears to continually increase through perihelion and beyond. The CO to $H_2O$ abundance ratio therefore increases throughout the plotted period and at the time of the DIF measurements is at ~60%. Unfortunately, the HRI-IR instrument is the only one that directly measured Garradd's $CO_2$ production rate, so a long temporal trend for $CO_2$ is not available for comparison.

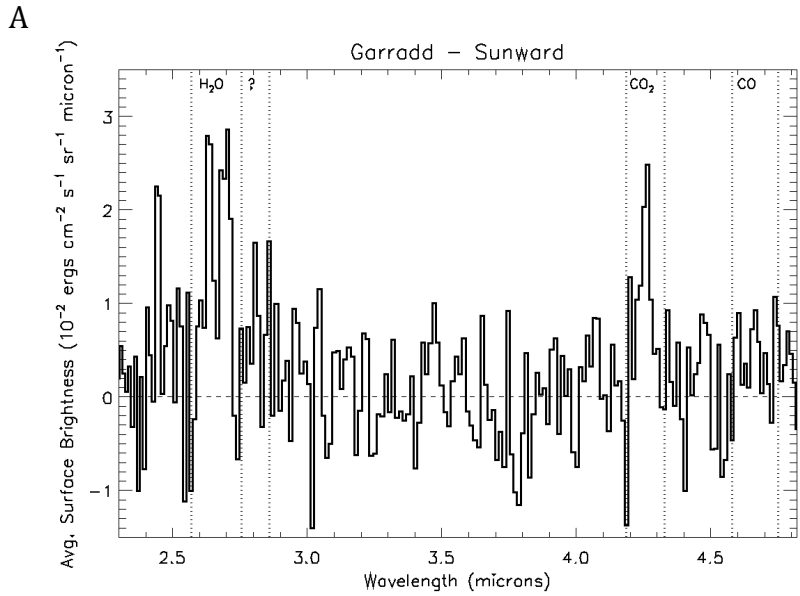

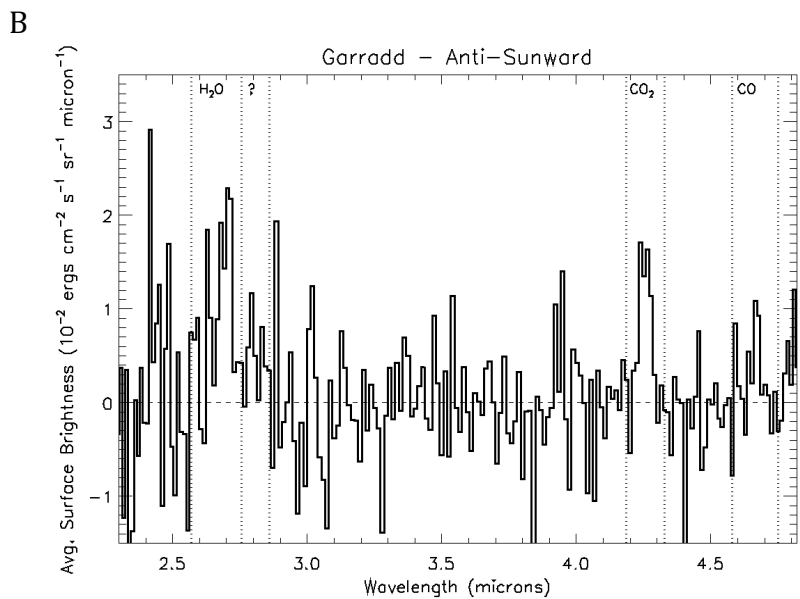

Figure 10 – Spectra of comet Garradd extracted from 2.3 to 4.82 μm along the projected Sun-comet line to investigate the sunward (A) and anti-sunward (B) asymmetry, especially in the $H_2O$ emission band. The radial extent in either direction is 8,420 km. $H_2O$ and $CO_2$ are more intense in the sunward direction as compared to the anti-sunward direction although the enhancements are within the uncertainties of the data (Table 5). The sunward spectrum of CO is dominated by noise whereas there is a peak in the anti-sunward spectrum, but the error bars of the average band surface brightnesses overlap. Thus, the HRI-IR data suggest but do not convincingly show a pronounced sunward vs. anti-sunward asymmetry, with H$_2$O and CO$_2$ enhanced in the sunward direction and CO more pronounced in the anti-sunward direction.

Table 1 – HRI-IR Observational Parameters.

| UTC of Observation | Days Post-Perihelion | Heliocentric Distance [AU] | Phase Angle [deg] | S/C Distance [AU] | Pixel Scale (along slit x along scan) [km] |
|---|---|---|---|---|---|
| 2012-03-26T07:33:44 -- 2012-03-26T23:21:56 | 94 | 2 | 36.4 | 1.44 | 2160 x 4320 |
| 2012-04-02T08:18:29 -- 2012-04-03T00:06:41 | 101 | 2.06 | 34.0 | 1.37 | 2050 x 4100 |

Table 2 – Scan Details.

| HRI-IR Scan Details | |
|---|---|
| Scans Per Day | 64 |
| Integration Time | 12 s per frame |
| Number of Frames | 17 per scan |
| Slit Width | 2 pixels per frame |
| Cadence of Scans | 1 scan per 15 min |

Table 3 – Garradd's dominant volatile inventory as observed by the DIF HRI-IR spectrometer.

| Parent Molecule | Effective Aperture Radius [km] | Average Surface Brightness [ergs cm$^{-2}$ s$^{-1}$ sr$^{-1}$] | Solar g-factor (@1 AU) [s$^{-1}$] | Average Column Density [molecule cm$^{-2}$] | Production Rate [molecule s$^{-1}$] |
|---|---|---|---|---|---|
| H$_2$O | 1,680 | 0.0062 ± 0.0015 | 2.1e-4 | 2.1e15 ± 0.5e15 | 3.5e28 ± 0.9e28 |
|  | 6,500 | 0.0029 ± 0.0005 | 2.6e-4 | 7.8e14 ± 1.3e14 | 5.1e28 ± 0.9e28 |
|  | 11,270 | 0.0016 ± 0.0003 | 2.8e-4 | 4.1e14 ± 0.7e14 | 4.6e28 ± 0.8e28 |
| CO$_2$ | 1,680 | 0.0029 ± 0.0008 | 2.0e-3 | 1.6e14 ± 0.5e14 | 2.8e27 ± 0.8e27 |
|  | 6,500 | 0.0015 ± 0.0003 | 2.5e-3 | 6.5e13 ± 1.4e13 | 4.2e27 ± 0.9e27 |
|  | 11,270 | 0.0008 ± 0.0002 | 2.6e-3 | 3.5e13 ± 0.7e13 | 3.9e27 ± 0.7e27 |
| CO | 1,680 | 0.0024 ± 0.0008 | 1.8e-4 | 1.7e15 ± 0.6e15 | 2.8e28 ± 1.0e28 |
|  | 6,500 | 0.0008 ± 0.0003 | 2.1e-4 | 4.3e14 ± 1.5e14 | 2.8e28 ± 1.0e28 |
|  | 11,270 | 0.0005 ± 0.0001 | 2.2e-4 | 2.6e14 ± 0.7e14 | 2.9e28 ± 0.8e28 |
|  | Effective Annulus Radius [km] |  |  |  |  |
| H$_2$O | 840 | 0.0062 ± 0.0015 | 2.1e-4 | 2.1e15 ± 0.5e15 | 3.5e28 ± 0.9e28 |
|  | 4,100 | 0.0027 ± 0.0002 | 2.6e-4 | 7.2e14 ± 1.2e14 | 5.9e28 ± 0.9e28 |
|  | 8,900 | 0.0010 ± 0.0001 | 2.8e-4 | 2.5e14 ± 0.5e14 | 4.5e28 ± 0.9e28 |
| CO$_2$ | 840 | 0.0029 ± 0.0008 | 2.0e-3 | 1.6e14 ± 0.5e14 | 2.8e27 ± 0.8e27 |
|  | 4,100 | 0.0014 ± 0.0002 | 2.5e-3 | 6.1e13 ± 1.3e13 | 5.0e27 ± 1.1e27 |
|  | 8,900 | 0.0005 ± 0.0001 | 2.6e-3 | 2.0e13 ± 0.3e13 | 3.5e27 ± 0.6e27 |
| CO | 840 | 0.0024 ± 0.0008 | 1.8e-4 | 1.7e15 ± 0.6e15 | 2.8e28 ± 1.0e28 |
|  | 4,100 | 0.0006 ± 0.0002 | 2.1e-4 | 3.6e14 ± 1.3e14 | 3.0e28 ± 1.1e28 |
|  | 8,900 | 0.0003 ± 0.00005 | 2.2e-4 | 1.9e14 ± 0.4e14 | 3.3e28 ± 0.6e28 |

Table 4 – CO production rates derived for Garradd from optically thin observations acquired at the JCMT.

| Date | Beam Size [km] | $T_{rot}$ [K] | Expansion Velocity [km s$^{-1}$] | $Q_{CO}$ [molecule s$^{-1}$] |
|---|---|---|---|---|
| 2011 Jul 28-31 | 25,000 | 24 | 0.30 | 2.43 + 1.2 / − 0.7 × 10$^{27}$ |
| 2011 Sep 23-25 | 20,000 | 52 | 0.45 | 5.23 + 1.4 / − 1.1 × 10$^{27}$ |
| 2011 Oct 23-25 | 18,000 | 52 | 0.50 | 1.26 + 1.0 / − 0.6 × 10$^{28}$ |
| 2012 Jan 6-8 | 16,000 | 45 | 0.55 | 1.30 + 1.0 / − 0.4 × 10$^{28}$ |

Table 5 – Average surface brightnesses measured with the HRI-IR to examine sunward and anti-sunward asymmetries in Garradd.

| Parent Molecule | Direction | Average Surface Brightness [ergs cm$^{-2}$ s$^{-1}$ sr$^{-1}$] |
|---|---|---|
| H$_2$O | Sunward | 0.0023 ± 0.0004 |
|  | Anti-Sunward | 0.0018 ± 0.0003 |
| CO$_2$ | Sunward | 0.0013 ± 0.0005 |
|  | Anti-Sunward | 0.0008 ± 0.0002 |
| CO | Sunward | 0.0008 ± 0.0005 |
|  | Anti-Sunward | 0.0004 ± 0.0002 |